\documentclass{amsart}
\usepackage{amssymb}
\usepackage{amsfonts}
\usepackage{amsmath}
\usepackage{graphicx}%
\usepackage{lastpage}
\usepackage[legalpaper,bookmarks=true,colorlinks=true,linkcolor=blue,citecolor=blue]{hyperref}
\usepackage{fancyhdr}
\usepackage{color}
\usepackage[mathlines]{lineno}
\usepackage{lscape}
\usepackage{epsfig}
\usepackage{natbib}
\usepackage{float}

\newtheorem{theorem}{Theorem}
\theoremstyle{plain}

\newtheorem{definition}{Definition}

\newtheorem{remark}{Remark}

\numberwithin{equation}{section}

\begin{document}
\Large
\title[The Theil-like family of Inequality Measures]{A Theil-like Class of Inequality Measures, its Asymptotic Normality Theory and Applications}

\begin{abstract}In this paper, we consider  a coherent theory about the asymptotic representations for a family of inequality indices called Theil-Like Inequality Measures (\textit{TLIM}), within a Gaussian field. The theory uses the functional empirical process approach. We provide the finite-distribution and uniform asymptotic normality of the elements of the TLIM class in a unified approach rather than in a case by case one. The results are then applied to some UEMOA countries databases.\\

\bigskip\noindent 
\author{$^{\dag}$ Pape Djiby Mergane}
\author{$^{\dag\dag}$ Tchilabalo Abozou Kpanzou}
\author{$^{\dag\dag\dag}$ Diam Ba}
\author{$^{\dag\dag\dag\dag}$ Gane Samb Lo}

\noindent $^{\dag}$ Pape Djiby Mergane\\
LERSTAD, Gaston Berger University, Saint-Louis, S\'en\'egal.\\
Email : mergane@gmail.com.\\

\noindent $^{\dag\dag}$ Tchilabalo Abozou Kpanzou (corresponding author).\\
Kara University, Kara, Togo.\\
Email : kpanzout@gmail.com\\ 

\noindent $^{\dag\dag\dag}$ Diam Ba\\
LERSTAD, Gaston Berger University, Saint-Louis, S\'en\'egal.\\
Email : diamba79@gmail.com.\\

\noindent $^{\dag\dag\dag \dag}$ Gane Samb Lo.\\
LERSTAD, Gaston Berger University, Saint-Louis, S\'en\'egal (main affiliation).\newline
LSTA, Pierre and Marie Curie University, Paris VI, France.\newline
AUST - African University of Sciences and Technology, Abuja, Nigeria\\
gane-samb.lo@edu.ugb.sn, gslo@aust.edu.ng, ganesamblo@ganesamblo.net\\
Permanent address : 1178 Evanston Dr NW T3P 0J9,Calgary, Alberta, Canada.\\

\noindent \textbf{keywords and phrases}. Inequality measures; Asymptotic behaviour; Asymptotic representations; functional empirical proces.\\

\end{abstract}

\maketitle

\newpage
\noindent \textbf{R\'esum\'e}. (French) Dans cet article, nous pr\'esentons une th\'eorie coh\'erente de repr\'esentations asymptotiques d'une famille de mesures d'ing\'egalit\'e
d\'enomm\'ee \textit{TLIM} dans un champ gaussien pr\'ecis. Notre m\'ethode est fond\'ee sur le processus empirique fonctionnel. Nous tirons de la 
repr\'esentation asymptotique les limites en distribution des estimateurs plug-in des membres de la famille en dimension finie.
Les r\'esultats sont ensuite appliqu\'es \'a des donn\'ees issues des pays de l'UEMOA.\medskip

\section{Introduction} \label{sec1}

\noindent In this paper, we deal with a modern weak theory for some large class of inequality indices that, further, will allow to handle easy comparison studies with different kinds of statistics. \\

\noindent According to earlier economists, inequality indices are functional relations between the income and the economic welfare (see \cite{dalton1920}). This explains, among others, the wide variety of such indices in the literature (See, e.g., \cite {cfa80a, cfa80b,cfa00}).

\bigskip

\noindent Such statistics, of course, have been widely studied with respect to a great variety of interests, including statistical characterizations and asymptotic properties (
See \cite{duclosdav}, \cite{barrett}, for recent studies).

\bigskip

\noindent  Recently,  \cite{greselin09} provided a mathematical investigation of these indices in a modern setting including Vervaat processes, L-statistics and empirical processes.

\bigskip

\noindent Having in mind the necessity of comparing inequality measures with  different kind of statistics such as growth statistics, we aim at providing a coherent asymptotic  weak theory for some class of inequality measures. Indeed we propose the functional empirical process setting (see \cite{vaart}) which provide natural Gaussian field in which many statistics used in Economics may be represented in.

\bigskip

\noindent Our best achievement consists of the asymptotic representations for the elements of our class of  inequality measures, in terms of the above mentioned Gaussian field. The results are illustrated in data driven applications, on Senegalese data for instance.\\

\bigskip \noindent The class on which we focus here is a functional family of inequality measures which gathers various ones around the central Theil measure. This class named after the Theil-Like Inequality Measure (TLIM) will be the central point of our study. It includes the Generalized Entropy Measure, the Mean Logarithmic Deviation (\cite{cfa0203,theil,cfa80a}), the different inequality measures of \cite{atkinson}, \cite{champernowne}, \cite{kolm76}, and the divergence of \cite{renyi61}.

\bigskip \noindent This means that, here, we will not discuss other inequality statistics such as the Gini, the Generalized Gini, the S-Gini, the E-Gini (See  \cite{barrett}). Those statistics and similar ones will be treated in separate papers.

\bigskip 
\noindent Now we are going to introduce our family. For that, let  $X$ denote the income  (or expense) random variable related to a given population. We assume that $X$ and its independent observations are defined on the probability space $\left(\Omega, \mathcal{A}, \mathbb{P} \right)$ and take their values in on some interval $\mathcal{V}_X \subset \mathbb{R}^*_+$  and have common cumulative distribution function \textit{(cdf)}, $F(x)$, $x \in \mathcal{V}_X$. In this paper, we only use Lebesgue-Stieljes integrals and for any measurable function $\ell : \mathbb{R}\rightarrow \mathbb{R}$, we have, whenever it makes sense,

$$
\mathbb{E}(\ell(X))=\int \ell(x) \ dF(x) \equiv \int_{\mathcal{V}_X} \ell \ d\mathbb{P}_X,
$$

\bigskip \noindent where $\mathbb{P}_X=\mathbb{P}X^{-1}$ is the measure image of $\mathbb{P}$ by $X$, but is also Lebesgue-Stieljes probability measure characterized by: $\mathbb{P}_X(]a,b])=F(b)-F(a)$ for any $-\infty \leq a\leq b \leq +\infty$.\\

\noindent Now, consider a sample of $n \geq 1$ individuals or households of that population and observe their income $X_1,$ $X_2,$ $\cdots,$ $X_n$. We define the following family of inequality indices, indexed by $\phi=(\tau, h, h_1, h_2) \in \mathcal{P}_{0}$ as follows

\begin{equation}\label{ineq1}
T_n(\phi,X) =  \tau\left( \frac{1}{n} \sum_{j=1}^{n} \left( \frac{h\left(X_j\right)}{h_1\left(\mu_n\right)} - h_2\left(\mu_n\right)
\right)\right), \: h_1(\mu_n) \neq 0,\\
\end{equation}

\noindent where $\mu_n=\frac{1}{n}\sum_{j=1}^{n}X_j$ is the empirical mean while  $h(x)$, $h_1(y)$, $h_2(z),$ and $ \tau(t)$ are real and measurable functions of $x, y, z \in \mathcal{V}_X$ and $t \in \mathbb{R}$. The exact form of $\mathcal{P}_{0}$ is not important here, in opposite to the conditions on the functions $\tau$, $h$, $h_1$ and $h_2$ under which the results are valid. In a future paper on the uniform limits in $\phi$, that class will be crucial.

\bigskip 
\noindent We will see below that $T_{n}$ under specific hypotheses on $\tau,$ $h,$ $h_1,$ $h_2$ and $\mu_n$, converges  to the exact inequality measure 

\begin{equation}\label{ineq2}
T(\phi,X)= \tau\left(\frac {1}{h_1\left(\mu\right)}\, \int_{\mathcal{V}_X}\,h\left(x\right)\,dF(x) - h_2\left(\mu\right)\right), \: h_1(\mu) \neq 0,
\end{equation}
where $\mu=\mathbb{E}\left(X\right)$ is  the mathematical expectation of $X$ that we suppose finite here. We will come back later on the function classes $\mathcal{F}_1,$ $\mathcal{F}_2,$ $\mathcal{F}_3$ and $\mathcal{F}_4$ in which $h,$ $\tau,$ $h_1$ and $h_2$ are supposed to lie. 

\bigskip

\noindent Each measure of this Theil-like family has its own particular properties, that are derived from the combination of different concepts. One may mention the concept of welfare criteria (\cite{atkinson}, \cite{sen73}), that of the analogy with analysis of risks (\cite{harsanyi53}, \cite{harsanyi55}, \cite{rothschild}), that of the complaints approach (\cite{temkin93})  etc.
The Theil inequality itself finds all its interest in the information-theoretic idea following that of main components (Kullback 1959). It is based on the three following  axioms: Zero-valuation of certainty, Diminishing-valuation of probability, Additivity of independent events. A deep review of such of individual properties for a number inequality measures can be found in Cowell (\cite {cfa80a,cfa80b,cfa00}) for instance.\\

\noindent It is worth mentioning that the TLIM presented here, is rather a mathematical form gathering a number of different measures.

\bigskip \noindent The rest of our paper is organized as follows. In Section \ref{sec2}, we describe the TLIM family and show how the particular indices are derived from it. In Section \ref{sec3}, we briefly recall the functional empirical processes setting. In  section \ref{sec4}, we deal with the  asymptotic theory of the TLIM, state and describe our main results and demonstrate them. Section 
\ref{sec5} is devoted to datadriven applications. We finish by a conclusion in Section \ref{sec6}.

\bigskip

\section{Description of the  TLIM}\label{sec2}

\bigskip

\noindent This inequality measures mentioned above are derived from (\ref{ineq1}) with the particular values of the mesurable functions $\tau, h, h_1$ and $h_2$ as described below for all $s > 0.$

\bigskip

\subsection{Generalized Entropy}

$$
GE_{n,\alpha}\left(X\right) = \frac{1}{n \alpha \left(\alpha - 1\right)} \sum_{j=1}^{n} \left( \left( \frac{X_{j}}{\mu_n} \right)^{\alpha} - 1 \right);
$$

	$$ \alpha\neq 0,\, \alpha \neq 1,\;  \tau(s) = \frac{s-1}{\alpha\left(\alpha-1\right)},\,h(s)=h_1(s)=s^{\alpha},\; h_2(s)\equiv 0.$$

\subsection{Theil's measure }

$$ Th_n\left(X\right) = \frac{1}{n} \sum_{j=1}^{n} \frac{X_j}{\mu_n} \log{ \frac{X_j}{\mu_n}};$$

	$$ \tau(s) = s,\; h(s)= s\,\log(s),\; h_1(s)=s,\; h_2(s)= \log(s).$$

\subsection{Mean Logarithmic Deviation}

$$
MLD_n\left(X\right) = \frac{1}{n}\sum_{j=1}^{n}\log\left(\frac{X_j}{\mu_n}\right)^{-1};$$

	$$ \tau(s) = s,\; h(s)= h_2(s)=\log(s^{-1}),\; h_1(s)\equiv 1.$$
	
	\subsection{Atkinson's measure}
	
	$$Atk_{n,\alpha}\left(X\right) = 1 - \frac{1}{\mu_n}\left( \frac{1}{n}\sum_{j=1}^{n} X^{\alpha}_j\right)^{1/\alpha};$$
	
	$$\alpha < 1 \textrm{ and } \alpha\neq 0,\;  \tau(s)=1-s^{1/\alpha},\; h(s)=h_1(s)=s^{\alpha},\; h_2(s)\equiv 0.$$

\subsection{Champernowne's measure}

$$Ch_n\left(X\right) = 1 - \frac{1}{\exp\left(-\frac{1}{n}\sum_{j=1}^{n}\log\frac{X_j}{\mu_n} \right)};$$

	$$  \tau(s) = 1 - \exp\left(s\right),\; h(s)=h_2(s)=\log(s),\; h_1(s)\equiv 1.$$

\subsection{Kolm's measure}

$$
Ko_{n,\alpha}\left(X\right) = \log\left(\frac{1}{n}\sum_{j=1}^{n}\exp\left(-\alpha(X_j - \mu_n)  \right) \right)^{1/\alpha};$$

	$$ \alpha > 0,\;  \tau(s) = \frac{1}{\alpha}\log(s),\; h(s)= h_1(s)=\exp (- \alpha s),\; h_2(s)\equiv 0.$$

\subsection{Divergence of Renyi  }

$$DR_{n,\alpha}(X) = \frac{1}{\alpha - 1} \log\left( \frac{1}{n}\sum_{j=1}^{n} \left(\frac{X_j}{\mu_n} \right)^\alpha  \right);$$

$$
\alpha \in \mathbb{R}_{+}\backslash\left\{1\right\},\;  \tau(s) = \frac{1}{\alpha -1 }\log(s),\; h(s)= h_1(s)= s^ \alpha,\; h_2(s)\equiv 0.
$$


\bigskip
\section{The functional empirical process }\label{sec3}

Let $Z_1, Z_2,\ldots, Z_n$ be a sequence of independent and identically distributed random elements defined on the probability space $\left( \Omega, \mathcal{A}, \mathbb{P} \right),$
 with values in some metric space $\left(S,d\right).$ Given a collection $\mathcal{F}$ of measurable functions $f: S \rightarrow \mathbb{R}$ satisfying 

$$\sup_{f \in \mathcal{F}}\left| f(z) - \mathbb{P}(f)\right| < \infty, \; \textrm{for every } z,$$

\noindent where $\mathbb{P}(f) = \mathbb{E}\left(f\left(Z\right)\right)$ is the mathematical expectation of $f(Z)$, the functional empirical process  (\textit{FEP}) based on the $(Z_j)_{j=1, ..., n}$ and indexed by $\mathcal{F}$ is defined by:
$$
\forall f \in \mathcal{F}, \mathbb{G}_n\left(f\right) = \frac{1}{\sqrt{n}}\sum_{j=1}^{n}\left( f(Z_j) - \mathbb{P}(f)\right).
$$

\noindent This process is widely studied in \cite{vaart} for instance. It is readily derived from the real Law of Larges Numbers (\textit{LLN}) and the real Central Limit Theorem (\textit{CLT}) that $\mathbb{P}_n(f) =  \frac{1}{n}\sum_{j=1}^{n}f\left(Z_j\right) \rightarrow \mathbb{P}\left(f\right)\; a.s.$ and that  $\mathbb{G}_n\left(f\right) \rightarrow \mathcal{N}\left(0,\sigma^2_f\right)$, where 
\begin{equation}\label{eq1}
\sigma^2_f = \mathbb{P}\left(\left( f - \mathbb{P}\left(f\right)\right)^2\right) < \infty,
\end{equation}

\noindent whenever $\mathbb{E}\left(f(Z)^2\right) < \infty$.\\

\noindent When using the \textit{FEP}, we are often interested in uniform \textit{LLN}'s and weak limits of the \textit{FEP} considered as stochastic processes. This gives the so important results on Glivenko-Cantelli classes and Donsker ones. Let us define them here (for more details see \cite{vaart}).

\bigskip

\noindent Since we may deal with non measurable sequences of random elements, we generally use the outer almost sure convergence defined as follows:

\bigskip

\noindent a sequence $U_n$ converges outer almost surely to zero, denoted by $U_n \rightarrow 0\, as^*,$  whenever there is a measurable sequence of measurable random variables $V_n$ such that
\begin{enumerate}
	\item $\forall n,\: \left|U_n\right| \leq V_n,$ 
	\item $V_n  \longrightarrow 0 \; a.s.$
\end{enumerate}

\noindent The weak convergence generally holds in $\ell^{\infty}\left(\mathcal{F}\right),$  the space of all bounded real functions defined on $\mathcal{F},$ equipped with the supremum norm  $\left\|x\right\|_\mathcal{F} = \sup_{f\in\mathcal{F} }\left| x(f)\right|.$

\bigskip

\begin{definition}
$\mathcal{F} \subset L_1(\mathbb{P})$ is called a \textit{Glivenko-Cantelli class} for $\mathbb{P}$, if
$$\lim_{n\rightarrow \infty}\left\| \frac{1}{n}\sum_{j=1}^{n}\left( f(Z_j) - \mathbb{P}(f)\right)
\right\|_{\mathcal{F}} = 0 \,\textrm{a.s.}^*. $$ 

\end{definition}

\bigskip

\begin{definition}\label{def2}
$\mathcal{F} \subset L_2(\mathbb{P})$ is called a \textit{Donsker class} for $\mathbb{P}$, or $\mathbb{P}$\textit{-Donsker class} if $\left\{\mathbb{G}_n\left(f\right); f\, \in \mathcal{F}\right\}$ converges in $l^{\infty}\left(\mathcal{F}\right)$ to a centered Gaussian process  $\left\{\mathbb{G}\left(f\right); f\, \in \mathcal{F}\right\}$ with covariance function
$$\Gamma\left(f,g\right) = \int_{\mathbb{R}}
\left(f(z) - \mathbb{P}(f)\right)\left(g(z) - \mathbb{P}(g)\right)\,d\mathbb{P}_Z(z)\,;\,\forall f,g \in \mathcal{F}.$$
\end{definition}


\begin{remark} 
When $S=\mathbb{R}$ and $\mathcal{F} = \left\{f_t= \mathbf{1}_{(-\infty,t]}, t \in \mathbb{R} \right\}$, $\mathbb{G}_n$ is called real empirical process and is  denoted by $\alpha_n.$
\end{remark}

\noindent In this paper, we only use finite-dimensional forms of the FEP, that is $\left(\mathbb{G}_n\left(f_i\right), i=1, \ldots, k \right).$  And then, any family $\left\{f_i, i=1, \ldots, k\right\}$ of measurable functions satisfying (\ref{eq1}), is a Glivenko-Cantelli and a Donsker class, and hence 

$$
\left(\mathbb{G}_n\left(f_i\right), i=1, \ldots, k \right) \stackrel{d}{\rightarrow} \left(\mathbb{G}\left(f_1\right), \mathbb{G}\left(f_2\right), \ldots, \mathbb{G}\left(f_k\right)
\right),
$$

\bigskip \noindent where $\mathbb{G}$ is the Gaussian process, defined in Definition \ref{def2}. We will make use of the linearity property of both $\mathbb{G}_n$ and $\mathbb{G}$. Let $f_1, \ldots, f_k$ be measurable functions satisfying (\ref{eq1}) and $a_i \in \mathbb{R}, i=1, \ldots, k$, then 
\begin{equation*}
\sum_{j=1}^{k} a_j \mathbb{G}_n \left(f_j\right) = \mathbb{G}_n\left(\sum_{j=1}^{k}a_j f_j\right) \stackrel{d}{\rightarrow} \mathbb{G}\left(\sum_{j=1}^{k} a_j f_j \right).\\
 \end{equation*}

\bigskip \noindent The materials defined here, when used in a smart way, lead to a simple handling of the problem which is addressed here.


\section{Our results} \label{sec4}

\noindent Let us introduce some notation.

$$
B_{h,n}=\frac{1}{n}\sum_{j=1}^{n}h\left(X_{j}\right),\; B_h =\int_{\mathcal{V}_X} h(x)\,dF(x);
$$

$$K_\phi=\tau^{\prime}\left(\frac{B_h}{h_1(\mu)} - h_2(\mu)\right) \neq 0;$$

\noindent for all $x \in \mathcal{V}_X$, we define the following function  
$$F_\phi(x)=K_\phi\left(\frac{1}{h_1(\mu)} h(x) - \left(\frac{B_h\,h^{\prime}_1(\mu)}{h^2_1(\mu)} +h^{\prime}_2(\mu)  \right) I_d(x)\right)$$

\noindent with  $I_d(x)\,=\,x,$ and $\tau^{\prime}$ is the derivative of the function $\tau.$

\bigskip \noindent The following general condition will be assumed in all the paper:\\

\begin{itemize}
	\item[(\textbf{C})] $h_1$ is not null in a neighborhood of $\mu.$ 
\end{itemize}

\noindent Here are our main results.\\

\subsection{Pointwise asymptotic laws} \label{ssec1}

Consider the following hypotheses based on the functions $h,$ $\tau,$ $h_1,$ $h_2$. The \textit{A1.x} series concern the almost-sure limits and the \textit{A2.x} the asymptotic normality.

\bigskip

\begin{itemize}
	\item[(\textbf{A1.1})] $\; \mathbb{E}h(X)\, < \, \infty;$  
	\item[(\textbf{A1.2})] $\; \tau$ is a continuous function on $\mathcal{V}_X;$
	\item[(\textbf{A1.3})] for $i \in \left\{1, 2\right\},$  $h_i(\mu) < \infty$ and $h_i$ is continuous  on $\mathcal{V}_X.$
\end{itemize}

\bigskip

\begin{itemize}
	\item[(\textbf{A2.1})] $\mathbb{E}h^2(X) < \infty,$ $\mathbb{E}\left(X\,h(X)\right) < \infty;$
	\item[(\textbf{A2.2})] $\tau$ is continuously differentiable such that $\tau^\prime \neq 0;$
	\item[(\textbf{A2.3})] $\forall i \in \left\{1, 2\right\},$ $h_i(\mu) < \infty,$ $h_i$ is continuously differentiable at $\mu.$
\end{itemize}

\bigskip
\noindent We have :

\begin{theorem}\label{theo1}
Suppose that the  conditions $(\textbf{C}),$ $(\textbf{A1.1}),$ $(\textbf{A1.2})$ and $(\textbf{A1.3})$ are satisfied, then
$T_n$ converges almost surely to $T.$
\end{theorem}

\bigskip

\begin{theorem}\label{theo2}
Suppose that the conditions $(\textbf{C}),$ $(\textbf{A2.1}),$ $(\textbf{A2.2})$ and $(\textbf{A2.3})$ are satisfied, and $K_\phi$ is finite. Then

\noindent (a) we have the following asymptotic representation in the empirical functional process 
$$
\sqrt{n} \left(\tau(I_n) - \tau(I)\right) = \mathbb{G}_n\left(F_{\phi}\right) + o_{\mathbb{P}}(1), \ as \ n\rightarrow +\infty,
$$

\noindent where

$$
F_{\phi}=\tau^{\prime}\left(\frac{\mathcal{E}h(Y)}{h_1(\mu)}-h_2(\mu)\right)\left(\frac{1}{h_1(\mu)}\,h - \left( \frac{\mathcal{E}h(Y) \,h^{\prime}_1(\mu)}{h^2_1(\mu)} + h^{\prime}_2(\mu)\right)\, I_d\right)
$$

\bigskip \noindent (b) and we have the convergence in distribution, as $n$ tends to infinity, of $\sqrt{n}(T_n(\phi,X) -  T(\phi,X))$  to centered normal Gaussian law: 

$$
\sqrt{n}(T_n(\phi,X) -  T(\phi,X)) \rightsquigarrow  \mathcal{N}(0, \sigma^2_{\phi}),
$$ 

\noindent where 

\begin{eqnarray*}
\sigma^2_\phi &=& \int \left( F_{\phi}(x)-\int F_{\phi}(x) \ d\mathbb{P}_X(x)\right)^2 \ d\mathbb{P}_X(x)\\
&=&a_{\phi}^2\, \mathbb{E}\left( h(X)-\mathbb{E}h(X)\right)^2 + b_{\phi}^2\,\mathbb{E}\left( X-\mu\right)^2\\
&-& 2 a_{\phi} b_{\phi}\, \mathbb{E}\left( h(X)-\mathbb{E}h(X)\right)\left(X-\mu\right),
\end{eqnarray*}
	
\bigskip \noindent with
	
\begin{equation*}
a_{\phi} = \frac{K_\phi}{h_1(\mu)}\: \textrm{ and } b_{\phi} = K_\phi\left(\frac{B_h\,h^{\prime}_1(\mu)}{h^2_1(\mu)} +h^{\prime}_2(\mu)  \right) \label{ab}.
\end{equation*}
\end{theorem}

\bigskip \noindent \textbf{Remark}. The main result is the one given in Point (a). From it, Point (b) is deduced in a straightforward way.\\

\noindent The results above cover all the \textit{TLIM} class. They should be particularized for the practitioner who would pick one of the elements of that class for analyzing data. Here are then the details for each case.

\subsection{Particular cases for pointwise results} \label{ssec2} \ ~~~~ \

\noindent\textbf{a.} \textit{The Theil's measure}

\noindent The empirical form of Theil measure is defined as follows

$$
Th_n = \frac{1}{\mu_n}\,\frac{1}{n}\,\sum_{j=1}^{n}\, X_j\log X_j - \log \mu_n,
$$

$$
\forall s  > 0,\; \tau(s)\,=\,s ,\; h(s)\,=\, s\,\log(s),\; h_1(s) \,=\, s,\; h_2(s)\, =\, \log(s),
$$

\noindent Denote by

$$
Th = \frac{1}{\mu} \int_{\mathcal{V}_X}\,x\log x\,dF(x) - \log \mu 
$$

\bigskip

\noindent the continuous form of the Theil measure.

\bigskip

\noindent All these functions are continuous on $\mathcal{V}_X,$ then the assumptions defined above become for the \textit{a.s.} requires that $\mathbb{E} X\log X$ is finite and $0 < \mu < \infty$. As for the asymptotic normality, we need that

$$\mathbb{E}\left|X\right|^2,\, \mathbb{E}\left|X\log X \right|^2\, \mathbb{E}\left|X^2\log X \right|^2\, \textrm{ are finite}.$$

\noindent And we have $K_\phi=1,$ $B_h=\mathbb{E}\left(X\log X\right)$. We conclude that $$\sqrt{n}\left(Th_n - Th\right) \leadsto \mathcal{N}(0, \sigma^2_{Theil})  $$

\noindent with $$\sigma^2_{Theil} = \frac{\mathbb{E}\left(X\log X\right)^2 }{\mu^2} + \frac{\mathbb{E}X^2}{\mu^2} \left(\frac{B_h}{\mu} + 1 \right)^2 - \frac{2 \mathbb{E}\left(X^2\log X \right) }{\mu^2}\left(\frac{B_h}{\mu} + 1 \right) - 1.$$

\bigskip

\noindent\textbf{b.} \textit{The Mean Logarithmic Deviation}

\bigskip

\noindent Let $$MLD_n = \frac{1}{n} \sum_{j=1}^{n} \log X^{-1} - \log \mu_n^{-1} $$

\noindent be the empirical form of the Mean Logarithmic Deviation. Its theoretical form is given as folllows

$$
MLD = \int_{\mathcal{V}_X} \log x^{-1}\,dF(x) - \log \mu^{-1}.
$$
 
\noindent These specific functions are given by:

$$
\forall s > 0, \;\; \tau(s) = s,\;\; h(s) = h_2(s) = \log s^{-1},\;\; h_1(s) \equiv 1.
$$

\bigskip

\noindent The consistency requires that $\mathbb{E} \log X < \infty$ and that $0 < \mu < \infty$ while the normality is got when

$$\mathbb{E}\left|X\right|^2,\mathbb{E} \left|\log X\right|^2 \textrm{ and } \mathbb{E}\left|X\log X \right| \textrm{are finite}.$$

\bigskip 

\noindent In that case, we find easily that $K_\phi=1,$ $B_h=\mathbb{E}\log X^{-1}$ and 

$$
\sqrt{n}\left(MLD_n - MLD\right) \leadsto \mathcal{N}(0, \sigma^2_{MLD})
$$

\noindent where 

$$
\sigma^2_{MLD} = \frac{\mathbb{E}\left(X^2 \right) }{\mu^2} + \mathbb{E}\left(\log^2 X \right) - \frac{2}{\mu} \mathbb{E}\left(X\log X\right) - \left(B_h + 1 \right)^2.
$$

\bigskip

\noindent\textbf{c.} \textit{The Champernowne's measure}

\bigskip

\noindent In this case, the specific functions are given by:

$$  \tau(s) = 1 - \exp\left(s\right),\; h(s)=h_2(s)=\log(s),\; h_1(s)\equiv 1.$$

\noindent And, the various forms are:

$$Ch_n = 1 - {\exp\left(\frac{1}{n}\sum_{j=1}^{n}\log\frac{X_j}{\mu_n} \right)};$$

$$Ch = 1 - \exp\left( \int{\mathcal{V}_X} \log\frac{x}{\mu}\,dF(x)  \right) .$$

\bigskip

\noindent We find that $Ch_n = \tau\left(- MLD_n\right)$ and $Ch = \tau\left(- MLD\right),$ where $MLD$ is the Mean Logarithmic Deviation.
 As $\tau$ is continuous on $\mathcal{V}_X,$ we consider the same hypotheses as in the case of Mean Logarithmic Deviation.\\ 
The function $\tau$ is continuously differentiable, we put $B_h = \mathbb{E}\log X^{-1}$ and $K_\phi = \frac{\exp(-B_h)}{\mu},$ then we have

$$
\sqrt{n}\left(Ch_n - Ch \right) \leadsto \mathcal{N}(0, \sigma^2_{Ch}) 
$$

\noindent with

$$
\sigma^2_{Ch} = K^2_\phi\,\sigma^2_{MLD}.
$$

\bigskip \noindent\textbf{d.} \textit{Cases of the Generalized Entropy $(\alpha\neq 0,\, \alpha \neq 1);$ the Atkinson's measure $(\alpha < 1 , \alpha\neq 0);$ the Divergence of Renyi $(\alpha > 0, \alpha \neq 1) .$}

\bigskip

\noindent We may gather these indices into one subclass by giving different values to the function $\tau$ and to the parameter $\alpha$, with this common expression

$$\forall s > 0, h(s) = h_1(s) = s^\alpha \textrm{ and } h_2 \equiv 0,$$ 

\bigskip

\noindent and then give a general description of the results. For that,  Let $I_{n,\alpha} = \mathbb{P}_{n}(h) / h_{1}(\mu_{n})$ and $ I_\alpha = \int_{\mathcal{V}_X} \frac{h(x)}{h_1(\mu)}\,dF(x).$
 
\bigskip

\noindent We require for consistency that $\mathbb{E}X^{\alpha} < \infty$ and that $\mu \neq 0$ and, for asymptotic normality that  

$$\mathbb{E}\left|X\right|^{2\alpha} < \infty,\; \mathbb{E}\left|X\right|^{2} < \infty \textrm{ and } \mathbb{E}\left|X\right|^{\alpha+1} < \infty.$$

\noindent Further, let $B_h= \mathbb{E}X^{\alpha}$ and $K_\phi = \tau^{\prime}(I_{\alpha})$. Then we get

$$
\sqrt{n}\left(I_{n,\alpha} - I_{\alpha}\right) = \mathbb{G}_n\left( \frac{h}{\mu^{\alpha}} - \frac{\alpha \mathbb{E}X^{\alpha}}{\mu^{\alpha}+1  } I_d\right) + o^*_{\mathbb{P}}(1),
$$

\noindent which tends towards a centered Gaussian process with variance

\begin{equation}\label{sigma2I}
\sigma^2_{I_\alpha} = \frac{1}{\mu^{2 \alpha}} \left( \mathbb{E}X^{2 \alpha} + \frac{\left(\alpha B_h \right)^2}{\mu^2} \mathbb{E}X^{2} - \frac{2 \alpha B_h}{\mu} \mathbb{E}X^{\alpha+1}
\right) - \frac{ B^2_h}{\mu^{2\alpha}}\left(1 - \alpha\right)^2.
\end{equation}

\bigskip 
\noindent Now, we may return to the individual cases. 

\noindent\textbf{d.1.} \textit{Generalized Entropy}

\noindent We find $K_\phi=1/(\alpha(\alpha - 1))$,  from there, we get the variance

$$
\sigma^2_{GE_\alpha} = K^2_\phi \sigma^2_{I_\alpha}, \textrm{where } \sigma^2_{I_\alpha} \textrm{ is given in Equation (\ref{sigma2I})}.
$$

\bigskip

\noindent\textbf{d.2.} \textit{Atkinson's measure }

\noindent  Put $K_\phi = \left(\mathbb{E}X^{\alpha}\right)^{(1/\alpha - 1)}/\alpha$. We similarly get that

$$
\sigma^2_{Atk_\alpha} = \frac{1}{\alpha^2}\left(\mathbb{E}X^\alpha \right)^{\left(\frac{1-\alpha}{\alpha}\right)^2}\, \sigma^2_{I_\alpha}.
$$

\bigskip

\noindent\textbf{d.3.} \textit{Divergence of Renyi}

\noindent By taking $K_\phi = \left( (\alpha - 1)\mathbb{E}X^{\alpha} \right)^{-1},$ we obtain by the same way, that

$$
\sigma^2_{DR_\alpha} = \frac{\sigma^2_{I_\alpha}}{\left((\alpha-1)\, \mathbb{E}X^\alpha\right)^2}
$$

\noindent where $\sigma^2_{I_\alpha}$ is given in (\ref{sigma2I}).

\bigskip

\noindent\textbf{e.} \textit{ Case of the Kolm's measure}

\bigskip

\noindent This index is defined for $\alpha > 0,$ and its specific functions are:

$$
\tau(s) = \frac{1}{\alpha}\log(s),\; h(s)= h_1(s)=\exp (- \alpha s),\; h_2(s)\equiv 0,\; \forall s > 0.
$$

\bigskip

\noindent Its empirical form is given by

$$
Ko_{n,\alpha} = \log\left(\frac{1}{n}\sum_{j=1}^{n}\exp\left(-\alpha(X_j - \mu_n)  \right) \right)^{1/\alpha};
$$

\bigskip

\noindent and its theoretical form is defined as follows 

$$
Ko_\alpha = \log\left(\int_{\mathcal{V}_X} \left(\frac{e^{-x}}{e^{-\mu}}\right)^\alpha dF(x)  \right)^{1/\alpha}.
$$

\bigskip

\noindent We need for consistency that $\mu < \infty$ and that $\mathbb{E}\exp(-\alpha X) < \infty$ and, for asymptotic normality that

\bigskip

$\mathbb{E}\left(\left|X\right|^2\right),$ $\mathbb{E}\left(\left|e^{-\alpha X}\right|\right),$ $\mathbb{E}\left(\left|e^{-2 \alpha X}\right|\right)$ and $\mathbb{E}\left(\left|X e^{-\alpha X}\right|\right)$  are finite.\\

\noindent Then we have  $B_h = \mathbb{E}\left(e^{-\alpha X}\right)$ and $K_\phi = \left({\alpha B_h e^{\alpha \mu}}\right)^{-1}.$

\bigskip

\noindent Put
$$
I_{n,\alpha} = \frac{1}{e^{-\alpha \mu_n}} \frac{1}{n} \sum_{j=1}^{n} e^{- \alpha X_j} \textrm{ and } I_\alpha = \frac{1}{e^{-\alpha \mu}} \int_{\mathcal{V}_X} e^{-\alpha x}\,dF(x).
$$

\noindent Then

$$
\sqrt{n}\left( I_{n,\alpha} - I_{\alpha}\right) = \mathbb{G}_n\left( e^{\alpha \mu}\left( h  + \alpha B_h I_d \right)  \right) + o^*_{\mathbb{P}}(1).
$$

\bigskip \noindent Since $Ko_{n,\alpha} = \tau(I_{n,\alpha}),$ we deduce that

$$
\sigma^2_{Ko_\alpha} = \frac{\mathbb{E}e^{- 2 \alpha X} }{\left(\alpha B_h \right)^2} + \mathbb{E}X^2 +  \frac{2}{\alpha B_h} \mathbb{E} \left(X e^{\alpha X}\right)  - \left(\frac{1}{\alpha} + \mu \right)^2.
$$

\noindent Finally, we summarize the used abbreviations in Table \ref{notations}, and, for each index, the expression of the function $F_\phi$ and $\mathbb{P}(F_\phi)$ in Table \ref{SummaryF} where we can find the expressions of $a_\phi$ and $b_\phi.$

\bigskip
\begin{center}
\begin{table}[h!]
\begin{tabular}{c|c}
\hline
Notations & Indices\\
\hline
\hline
$GE(\alpha),$ $\alpha \neq 0, 1$ & Generalized Entropy with parameter $\alpha$ \\
\hline
THEIL & Theil \\
\hline
MLD & Mean Logarithmic Deviation\\
\hline
ATK$(\alpha),$ $\alpha < 1$ and $\alpha \neq 0$ & Atkinson with parameter $\alpha$ \\
\hline
CHAMP & Champernowne \\
\hline
KOLM$(\alpha) \alpha> 0$& Kolm with parameter $\alpha$\\
\hline
DR$(\alpha) \alpha \geq 0, \alpha \neq 1$& Divergence of Renyi with parameter $\alpha$\\
\hline
\hline
\end{tabular}
\caption{Notations of the indices} \label{notations} 
\end{table}
\end{center}

\bigskip

\begin{center}
\begin{table}[h!]
\begin{tabular}{c|c|c|c}
\hline
Indices & $B_h$ & $F_\phi(x),\; \forall x\in \mathcal{V}_{X}$ & $\mathbb{P}(F_\phi)$\\
\hline
\hline
  & & \\
$GE(\alpha)$ & $\int_{\mathcal{V}_X} x^{\alpha}\,dF(x)$ & $\frac{1}{\alpha(\alpha-1)}\frac{1}{\mu^\alpha}\left(x^\alpha - \frac{\alpha B_h}{\mu} x\right) $ & $\frac{- B_h}{\alpha \mu^\alpha}$ \\
  & & \\
\hline
  & & \\
THEIL & $\int_{\mathcal{V}_X} x\log x\,dF(x)$ & $\frac{1}{\mu}\left( x\log x -\left(\frac{B_h}{\mu} + 1 \right)x\right)$& $- 1$\\
  & & \\
\hline
  & & \\
MLD & $\int_{\mathcal{V}_X} \log x^{-1}\,dF(x)$  & $\frac{1}{\mu}x - \log x$ & $1 + B_h$\\
  & & \\
\hline
  & & \\
ATK$(\alpha)$& $\int_{\mathcal{V}_X} x^{\alpha}\,dF(x)$ & $\frac{B^{1/\alpha}_h}{\mu}\left(\frac{1}{\mu}x - \frac{B^{-1}_h}{\alpha}x^\alpha  \right)$ & $\left(1 - \frac{1}{\alpha}\right) \frac{B^{1/ \alpha}_h}{\mu}$ \\
  & & \\
\hline
  & & \\
CHAMP & $\int_{\mathcal{V}_X} \log x\,dF(x)$ & $\left(\frac{1}{\mu}x-\log x \right)\frac{\exp(B_h)}{\mu}$ & $\frac{1 - B_h}{\mu} \exp(B_h)$ \\
  & & \\
\hline
  & & \\
KOLM$(\alpha)$& $\int_{\mathcal{V}_X}\exp\left(-\alpha x\right) \,dF(x) $ & $x+\frac{1}{\alpha B_h}\exp{\left(- \alpha x\right)}$& $\mu + \frac{1}{\alpha}$\\
 & & \\
\hline
  & & \\
DR$(\alpha)$ & $\int_{\mathcal{V}_X} x^{\alpha}\,dF(x)$ & $\frac{1}{\alpha-1}\left(\frac{1}{B_h}x^{\alpha} - \frac{\alpha}{\mu}x\right) $ & $-1$ \\
  & & \\
  & & \\
\hline
\hline    
\end{tabular}
\caption{Summary of the functions $F$ for each index} \label{SummaryF} 
\end{table}
\end{center}

\bigskip

\section{Proof of Theorems \ref{theo1} and \ref{theo2}} \label{sec5}

\noindent \textbf{Proof of Theorem \ref{theo1}}.\\

\noindent On one hand, denote by 

\begin{equation}\label{I}
I_n = \frac{\mathbb{P}_{n}(h)}{h_1(\mu_n)} - h_2(\mu_n)  \textrm{ and } I = \frac{\mathbb{P}(h)}{h_1(\mu)} - h_2(\mu),\\
\end{equation}

\noindent by decomposing the difference of  $I_n$ and $I,$ we get the next equality

$$
I_n - I = \frac{\left(\mathbb{P}_n - \mathbb{P}\right)(h) }{h_1(\mu_n)} - \frac{\mathbb{P}(h)}{h_1(\mu)h_1(\mu_n)}\left( h_1(\mu_n) - h_1(\mu) \right) - \left( h_2(\mu_n) - h_2(\mu)\right).$$

\bigskip

\noindent As for all $i=1, 2;$ the function $h_i$ is continuous  on $\mathcal{V}_X$ and using the fact that $\mu_n$ converges almost surely to $\mu,$ then we have when $n$ tends to infinity

\begin{equation}\label{eq1consistency}
h_i(\mu_n) \stackrel{as}{\longrightarrow} h_i(\mu) < \infty. 
\end{equation}

\bigskip

\noindent We have also

$$
\left(\mathbb{P}_n - \mathbb{P}\right)\left(h\right) = \frac{1}{n} \sum_{j=1}^{n}\left(h(X_j) - \mathbb{E}h(X_j) \right).
$$

\noindent Or the sequence of the random variables $\left\{h(X_j)\right\}_{j=1, \cdots, n}$ is independent and identically distributed, and as $\mathbb{E}h(X) < \infty $ by the hypothesis $(\textbf{A1.1}),$ then the Law of Large Numbers implies that

\begin{equation}\label{eq2consistency}
\left(\mathbb{P}_n - \mathbb{P}\right)\left(h\right)  \stackrel{as}{\longrightarrow} 0.
\end{equation}

\noindent Finally, using (\ref{eq1consistency}) and  (\ref{eq2consistency}), we get 
$$
I_n \stackrel{as}{\longrightarrow} I, \textrm{ when } n \rightarrow \infty.
$$

\bigskip

\noindent On the other hand, as $\tau$ satisfies the hypothesis $(\textbf{A1.2}),$ then we deduce that
$$
T_n \stackrel{as}{\longrightarrow} T, \textrm{ when } n \rightarrow \infty. \ \square
$$

\bigskip \noindent \textbf{Proof of Theorem \ref{theo2}}.\\

\noindent Using the equation (\ref{I}), we have

$$
I_n - I = \frac{\left(\mathbb{P}_n - \mathbb{P}\right)(h) }{h_1(\mu_n)} - \frac{B_h}{h_1(\mu)h_1(\mu_n)}\left( h_1(\mu_n) - h_1(\mu) \right) - \left( h_2(\mu_n) - h_2(\mu)\right).$$

\bigskip

\noindent  Since  $h_i$ is continuously differentiable at $\mu$  for $i = 1, 2,$ we get

\bigskip

$$
h_i(\mu_n) - h_i(\mu)  = h^\prime_i(\mu)\, \left(\mathbb{P}_n - \mathbb{P}\right)(I_d) + o_{\mathbb{P}}(n^{-\frac{1}{2}}).
$$

\bigskip \noindent Then

$$
I_n - I =  \frac{\left(\mathbb{P}_n- \mathbb{P}\right)(h) }{h_1(\mu_n)} 
 - \frac{B_h}{h_1(\mu) \, h_1(\mu_n)}\left(h^{\prime}_1(\mu)\, \left(\mathbb{P}_n - \mathbb{P}\right)\left(I_d\right) + o^{*}_{\mathbb{P}}(n^{-\frac{1}{2}}) \right)
$$
$$ - h^{\prime}_2(\mu)\, \left(\mathbb{P}_n - \mathbb{P}\right)\left(I_d\right)
  + o^{*}_{\mathbb{P}}(n^{-\frac{1}{2}}).
$$

\noindent But 

$$
\frac{B_h}{h_1(\mu) \, h_1(\mu_n)}  o_{\mathbb{P}}(n^{-\frac{1}{2}})   + o^{*}_{\mathbb{P}}(n^{-\frac{1}{2}}) =  o_{\mathbb{P}}(n^{-\frac{1}{2}}),
$$

\bigskip

\noindent then we get the next expression 

$$
I_n - I =  \frac{\left(\mathbb{P}_n - \mathbb{P}\right)\left(h\right)}{h_1(\mu)}  - \left( \frac{B_h\,h^{\prime}_1(\mu)}{h^2_1(\mu)}\, + h^{\prime}_2(\mu) \right)\, \left(\mathbb{P}_n - \mathbb{P}\right)\left(I_d\right)  + o_{\mathbb{P}}(n^{-\frac{1}{2}}).
$$

\bigskip \noindent Then

$$
\sqrt{n}(I_n - I) =  \frac{1}{h_1(\mu)} \mathbb{G}_n\left(h\right) 
 -  \left( \frac{B_h\,h^{\prime}_1(\mu)}{h^2_1(\mu)}\, + h^{\prime}_2(\mu) \right)\, \mathbb{G}_n\left(I_d\right)
  +  o_{\mathbb{P}}(1).
$$

\bigskip

\noindent By the linearity property of $\mathbb{G}_n$, we get

$$
\sqrt{n}\left(I_n - I\right) = \mathbb{G}_n\left(\frac{1}{h_1(\mu)}\,h - \left( \frac{B_h\,h^{\prime}_1(\mu)}{h^2_1(\mu)} + h^{\prime}_2(\mu)\right)\, I_d\right) + o_{\mathbb{P}}(1).
$$

\bigskip

\noindent Since $K_\phi$ is finite by assumption,  we apply a gain the delta-method to the function $\tau$ to have

$$
\sqrt{n} \left(\tau(I_n) - \tau(I)\right) = \mathbb{G}_n\left(K_\phi\left(\frac{1}{h_1(\mu)}\,h - \left( \frac{B_h\,h^{\prime}_1(\mu)}{h^2_1(\mu)} + h^{\prime}_2(\mu)\right)\, I_d\right)
\right) + o_{\mathbb{P}}(1).
$$

\bigskip

\noindent Using the notations of the equation (\ref{ab}), therefore

$$
\sqrt{n} \left(\tau(I_n) - \tau(I)\right) = \mathbb{G}_n(a_\phi h - b_\phi I_d) +  o^{*}_{\mathbb{P}}(1) = \mathbb{G}_n(F_\phi) +  o_{\mathbb{P}}(1)$$

\noindent and we easily obtain by ~\eqref{eq1} the variance $\sigma^2_{\phi}$. This ends the proof of Theorem \ref{theo2}. $\blacksquare$\\

\section{Data driven applications and variance computations}\label{sec6}
We here give data driven applications to show how our results work. We consider the ESAM2 (Enqu\^ete S\'en\'egalaise aupr\`es des M\'enages, $2^{\textrm{\`eme}}$ \'edition) and the ESPS (Enqu\^ete de Suivi de la Pauvret\'e au S\'en\'egal) databases respectively collected in 2001-2002 and in 2005-2006. (See \cite{ansd}). S\'en\'egal is a member of UEMOA. In both databases, we consider expense variables aggregated at the level of Heads households as indicators of welfare.

\bigskip

\noindent We present the data in Table \ref{datadescriptive}.
\small
\begin{center}

\begin{table}[h]
\begin{tabular}{|l||c|c|c|}
\hline
  Data    & Years of data collection &  Number of households & Mean of the expenses\\
\hline
\hline
ESAM2 & 2001-2002 &  $6565$ & $995.20$\\
\hline
ESPS & 2005-2006 &  $13568$ & $898.70$\\
\hline
\end{tabular}

\bigskip

\caption{Descriptive Statistics for the Distribution} \label{datadescriptive} 
\end{table}
\end{center}

\bigskip
\noindent We proceeded to the computations of the inequality measures and the corresponding variances using \cite{R}. We obtained the results in Table \ref{variances}.
\small
\begin{center}

\begin{table}[h!]
\begin{minipage}[t]{.3\linewidth}
    \begin{tabular}{|l||c|c|}
    \hline
    \multicolumn{3}{|c|}{ESAM2} \\
    \hline
TLIM  & $ T(\textrm{ in }\% )$ & $\sigma^2_{\phi}$  \\
\hline
\hline
$GE(.5)$ & $36.362$ &  $1.643$  \\
\hline
$GE(2)$ & $100.984$ &  $148.274$ \\
\hline
THEIL & $43.102$ &  $4.371$ \\
\hline
MLD & $34.286$ & $1.024$ \\
\hline
ATK$(.5)$ &  $17.355$& $0.339$ \\
\hline
ATK$(-.5)$ & $37.497$ &  $0.532$ \\
\hline
CHAMP & $4.846$ & $1.636$ \\
\hline
DR$(.5)$ & $19.061$ & $0.339$ \\
\hline
DR$(2)$ & $110.515$ & $65.043$ \\
\hline
    \end{tabular}
\end{minipage}
\hfil
\begin{minipage}[t]{.3\linewidth}
    \begin{tabular}{|l||c|c|}
    \hline
    \multicolumn{3}{|c|}{ESPS} \\
    \hline
TLIM & $ T(\textrm{ in }\% )$ & $\sigma^2_{\phi}$  \\
\hline
\hline
$GE(.5)$ & $22.684$ &  $0.238$  \\
\hline
$GE(2)$ & $34.206$ &  $3.709$ \\
\hline
THEIL & $24.007$ &  $0.411$ \\
\hline
MLD & $23.060$ & $0.202$ \\
\hline
ATK$(.5)$ &  $ 11.021$& $0.053$ \\
\hline
ATK$(-.5)$ & $29.591$ &  $0.280$ \\
\hline
CHAMP & $3.334$ & $0.519$ \\
\hline
DR$(.5)$ & $11.677$ & $0.053$ \\
\hline
DR$(2)$ & $52.124$ & $5.231$ \\
\hline
    \end{tabular}
\end{minipage}
\bigskip
\caption{Results of the variances computations } \label{variances} 
\end{table}
\end{center}

\section{Conclusion}\label{sec6}

\noindent The family we introduced allows a flexible and unified approach in the asymptotic theory of a class of inequality indices. In parallel, the computer packages also may be presented in more compact forms. We illustrated both aspects (theoretical and computational) in the paper. Hence the practitioner has all he needs about these indices in one place. But we only studied the finite dimensional limits. In a future paper, we will try to present uniform asymptotic laws of the family index by the parameter $\phi=(\tau, h, h_1, h_2)$.

\end{document}